\documentclass[aps,prl,twocolumn,showpacs,preprintnumbers,amsmath,amssymb,floatfix,nofootinbib]{revtex4-1}
\usepackage{graphicx}
\usepackage{color}
\usepackage{colordvi}

\usepackage[normalem]{ulem}

\newcommand{\beqn}{\begin{eqnarray}}
\newcommand{\eeqn}{\end{eqnarray}}
\newcommand{\be}{\begin{equation}}
\newcommand{\ee}{\end{equation}}

\newcommand{\ba}{\begin{array}{c}}
\newcommand{\bat}{\begin{array}{cc}}
\newcommand{\ea}{\end{array}}

\newcommand{\bi}{\begin{itemize}}
\newcommand{\ei}{\end{itemize}}

\newcommand{\Frac}[2]{\frac{\displaystyle #1}{\displaystyle #2}}

\newcommand{\cO}{{\cal O}}

\newcommand{\mS}{\mathcal{S}}
\newcommand{\mT}{\mathcal{T}}

\newcommand{\lsim}{\stackrel{<}{_\sim}}
\newcommand{\gsim}{\stackrel{>}{_\sim}}

\newcommand{\Int}{\displaystyle{\int}}

\newcommand{\bear}{\begin{eqnarray}}
\newcommand{\eear}{\end{eqnarray}}
\newcommand{\nn}{\nonumber}

\begin{document}


\title{
A broad 
diphoton resonance at the TeV?
Not alone
}
\author{Pablo Roig $^{1}$}
\author{Juan Jos\'e Sanz-Cillero $^{2}$}
\affiliation{$^{1}$
Departamento de F\'isica, Centro de Investigaci\'on y de Estudios Avanzados del Instituto Polit\'ecnico Nacional,
Apartado Postal 14-740, 07000 M\'exico City, M\'exico}

\affiliation{${}^2$
Departamento de F\'\i sica Te\'orica I, Universidad Complutense de Madrid, E-28040 Madrid, Spain}

\begin{abstract}

The hint for a possible resonance in the diphoton channel with mass of 750~GeV disappeared in the data presented at ICHEP'16 by ATLAS and CMS.
However, the diphoton final state remains as one of the golden channels for new physics discoveries at the TeV scale in the LHC experiments. This motivates us to analyze
model-independently the implications of an $\cal{O}$(TeV) bump in the $\gamma\gamma$ final state.
By means of forward sum-rules for $\gamma\gamma$  
scattering, we show that
a spin--0 resonance with mass of the order of the TeV and
a sizable $\gamma\gamma$  
partial width
--of the order of a few GeV--
must be accompanied by higher spin resonances with $J_R\geq 2$
with similar properties,
as expected in strongly coupled extensions of the Standard Model
or, alternatively, in
higher dimensional deconstructed duals.
Furthermore, independently of whether the putative $\cal{O}$(TeV) candidate is a scalar or a tensor, the
large contribution to the forward sum-rules in the referred scenario
implies the presence of
states in the spectrum with $J_R\geq 2$,
these high spin particles being
a manifestation of new extra-dimensions or composite states of a new strong sector.
\end{abstract}

\maketitle

\section{Introduction}

Last December both ATLAS and CMS came up with significant local excesses in the diphoton
spectrum at 750~GeV of more than $3\,\sigma$ and a global significance in the range of 1--2 $\sigma$~\cite{exp}.
However, the results presented by both collaborations at the ICHEP'16 conference
in Chicago \cite{ICHEP16} suggest that
earlier data were a statistical fluctuation. Still, since the di-photon final state
remains to be one of the most promising channels in new physics searches,
we consider it might be interesting to point out
some model-independent implications deriving from having
a resonance with large $\gamma\gamma$ 
partial width
, of the order of a few GeV.~\footnote{See e.g. Ref.~\cite{Staub:2016dxq} for a review on weakly-coupled scenarios giving
rise to a narrow-width  
resonance
with mass in the TeV range.}
Fully general quantum field-theory properties --analyticity, crossing symmetry and unitarity-- imply a series of sum-rules
for $\gamma\gamma$ scattering~\cite{photon-SR}  
. They require a precise cancellation between the different helicity contributions
to the scattering amplitude that --remarkably--
cannot be accomplished by just spin--0 intermediate exchanges.
Likewise, a large resonance contribution cannot be cancelled by a weakly interacting
background unless the non-resonant (non-R) loop diagrams get 
large and the theory abandons
the perturbative regime.
As a result of this, resonances with $J_R\geq 2$ must appear in this channel to fulfill
the sum-rules.
The exchange of states with spin $J_R\geq 2$ in the crossed channels leads to
partial-wave amplitudes that diverge like $s^{J_R-1}\ln{s}$ (see e.g.~\cite{PWA}).
This can only be
solved through the
infinite tower of resonances of increasing spin
{\it \`a la Regge} that appears in strongly-coupled theories like Quantum Chromodynamics (QCD) or
their equivalent duals --string theory scenarios
{\it \`a la Veneziano}~\cite{Veneziano} or holographic 
models~\cite{RS}--, these high spin particles being 
a manifestation of new extra-dimensions or composite states of a new strong sector~\cite{Sanz:2016auj}.

\section{Forward sum-rules}

We analyze the forward scattering amplitude
 ($t=0$),
$T_{\Delta\lambda}$, as a function of the variable $\nu\equiv (s-u)/2$,
\footnote{
The use of the kinematical variable $\nu\equiv (s-u)/2$ is customary in fixed-$t$ dispersive analyses
of scattering amplitudes with definite $s\leftrightarrow u$ crossing properties (see in general Ref.~\cite{PWA}).
In particular, for $t=0$ one has $\nu=s$.
}
\bear
T_{\Delta \lambda}(\nu) =
T(V(k,\lambda)V(k',\lambda')\to V(k,\lambda)V(k',\lambda')) \, ,\;\;
\eear
between real spin--1 particles $V$ with helicity difference $\Delta \lambda=|\lambda-\lambda'|$.
In particular, we will focus on the case 
with  $V=\gamma$  
 , i.e., the forward $\gamma\gamma$  
scattering.

Analyticity, crossing symmetry and unitarity imply the crossing and forward
once-subtracted dispersion relation
\bear
\label{eq.disp-rel}
T_{\Delta \lambda}(\nu) &=& T_{\overline{{\Delta \lambda}}}(-\nu)\,=\,    T_{\Delta \lambda}(0)
\\
&&\hspace*{-1.5cm}
 +\Frac{\nu}{\pi}\Int_{    \nu_{\rm th}      }^\infty \Frac{\rm d\nu'}{\nu'}
\left(\Frac{{\rm Im}T_{\Delta \lambda}(\nu'+i\epsilon)}{\nu'-\nu}
- \Frac{{\rm Im}T_{\overline{{\Delta \lambda}}} (\nu'+i\epsilon)}{\nu'+\nu}
\right) \, ,
\nn
\eear
  with the production threshold $\nu_{\rm th}$ and
$\overline{\Delta\lambda}\equiv 2 - \Delta \lambda$. 
The low-energy forward $\gamma\gamma$ scattering is provided by the Euler-Heisenberg effective field theory (EFT)~\cite{EH-EFT},
 \bear
T_{\Delta\lambda}(\nu)\,\, \stackrel{\nu\to 0}{=}\,\, \cO(\nu^2)\, . 
\label{eq.EFT}
\eear 
This same result applies to
other
unbroken gauge theories
(like, for instance, the scattering among same-color gluons,
$g^a g^a \to g^a g^a$, that we do not discuss here).
In the case of spontaneously broken theories it is possible to generate
dimension 4 operators in the low-energy EFT, as in
the forward $ZZ$ scattering in the Standard Model (SM) due to the tree-level Higgs ($h$) exchange
\cite{BEH...}. We will focus in the following on the light-by-light scattering.

Matching Eq.(\ref{eq.disp-rel}) and
the EFT result (\ref{eq.EFT})
up to $\cO(\nu)$
leads to the forward photon sum-rule
\bear
0 &=& \Frac{1}{\pi}\Int_{   \nu_{\rm th}      }^\infty \Frac{\rm d\nu'}{(\nu')^{2}}
\left( {\rm Im}T_2(\nu'+i\epsilon) -  {\rm Im}T_0 (\nu'+i\epsilon) \right) \, ,
\label{eq.SR}
\eear
equivalent to the Roy-Gerasimov-Moulin
sum-rule for the inclusive
$\gamma\gamma$ cross section
${    \sigma_{\Delta\lambda}=   \sigma(\gamma(k,\lambda)\gamma(k',\lambda')\to X)   }$~\cite{photon-SR}:
\bear
0 &=& \frac{1}{\pi}  \Int_{    \nu_{\rm th}    }^\infty \Frac{\rm d\nu'}{\nu'}
\left[ \sigma_2(\nu') -  \sigma_0 (\nu') \right] \, ,
\label{eq.RGM-SR}
\eear
by means of the relation $\sigma_{\Delta\lambda}(\nu')={\rm Im}T_{\Delta\lambda}(\nu'+i\epsilon)/\nu'$.

The resonant contribution to the spectral function
(for $\nu\geq 0$)
is given by
\bear
&&  {\rm Im}T_{\Delta\lambda}(\nu+i\epsilon) \bigg|_R
\\
&&\quad =  \sum_R 16 \pi^2 \, (2 J_R+1)
\, M_R\, \Gamma_{R\to [\gamma\gamma]_{\Delta\lambda}}\,\, \delta(\nu-M_R^2) \, ,
\nn
\eear
and turns (\ref{eq.SR}) into
    \bear
0 &=& \sum_R\, 16\pi \, (2J_R+1) \,
\Frac{ ( \Gamma_{R\to [\gamma\gamma]_2}   - \Gamma_{R\to [\gamma\gamma]_0} ) }{M_R^3}
\nn\\
&&\qquad  \mbox{+ non-R,}
\label{eq.SR-narrow}
\eear
where
$\Gamma_{R\to [\gamma\gamma]_0}/2=\Gamma_{R\to \gamma(+)\gamma(+)}=\Gamma_{R\to \gamma(-)\gamma(-)}$,
$\Gamma_{R\to [\gamma\gamma]_2}=\Gamma_{R\to \gamma(+)\gamma(-)}$
and $\Gamma_{R\to \gamma\gamma}= \Gamma_{R\to [\gamma\gamma]_0}+ \Gamma_{R\to [\gamma\gamma]_2}$.
The `non-R' term  
is provided by the contributions to the spectral function from
loop diagrams in $\gamma\gamma\to \gamma\gamma$ without
an intermediate $s$--channel resonance.

The importance of this sum-rule relies on the fact that the lowest-spin resonances
($J_R=0$)
only decay into $\gamma\gamma$  
states with $\Delta\lambda=0$
and hence give a negative contribution to the sum-rule (\ref{eq.SR-narrow}). The positive
terms with $\Gamma_{R\to [\gamma\gamma]_2}$
only appear for higher spin resonances with $J_R\geq 2$~\cite{photon-SR,Pascalutsa:2010,Pascalutsa:2012,Panico:2016}:
\bear
&&
\Gamma_{R\to \gamma\gamma}= \Gamma_{R\to [\gamma\gamma]_0}\, , \,\,\, \Gamma_{R\to [\gamma\gamma]_2}=0\quad
\mbox{ for } J_R=0 \, ,
\\
&& \Gamma_{R\to \gamma\gamma}= \Gamma_{R\to [\gamma\gamma]_2}\, , \,\,\, \Gamma_{R\to [\gamma\gamma]_0}=0\,\,\,
\mbox{ for } J_R=3,5,7...
\nn
\eear
Resonances with $J_R=1$ are forbidden by the Landau-Yang theorem~\cite{Landau-Yang}
and those with $J_R=2,4,6...$ can in principle decay
into $[\gamma\gamma]_0$ and $[\gamma\gamma]_2$ states~\cite{Panico:2016}.
In QCD
the $\gamma\gamma$ decay of the lowest-lying spin--2 resonances ($\mT=a_2,f_2,f_2'$)
predominantly occurs with helicity $\Delta\lambda=2$~\cite{Tensor-Helicity-2},
i.e., $\Gamma_{\mT \to \gamma\gamma}\approx  \Gamma_{\mT\to [\gamma\gamma]_2}$.
The sum-rule~(\ref{eq.SR}) is mostly saturated by
the lightest pseudoscalar ($\pi^0,\eta,\eta'$)
and tensor ($a_2,f_2,f_2'$) mesons~\cite{Pascalutsa:2010,Pascalutsa:2012},
with the large spin--2 positive contribution cancelling out to a large extent
the large negative spin--0 contribution.
This situation resembles the case of (spin-2) massive
gravitons $G$~\cite{RS,KK-graviton,Gouzevitch:2013qca},
where the decay $G\to V(\lambda)V(\lambda')$
always occurs with $\Delta\lambda=2$ as the graviton couples
to the stress-energy tensor of the gauge field $V=\gamma, g^a$.

 Notice that the sum-rule studied here only relies on unitarity, analyticity and crossing symmetry
and is therefore fulfilled in any possible beyond the SM (BSM) extension that assumes these general properties.

\section{Sum-rule with an \cal{O}(TeV)  
spin--0 resonance}

The hint for a possible diphoton resonance with mass 750~GeV in the 2015 \cite{exp} and early ($\sim3 fb^{-1}$) 2016 data \cite{Moriond} has not been confirmed analyzing the
$15.4\,(12.9) fb^{-1}$ data by the ATLAS (CMS) Collaboration. We wish, however, to illustrate how the sum rule Eq.~(\ref{eq.SR-narrow}) 
allows us 
to derive model independent constraints
which may be useful to check the consistency of a supposed isolated future bump
in the diphoton channel. 

As our final purpose is to show the need of resonances with spin $J_R\geq 2$ in the spectrum, 
we will assume from now on that 
an \cal{O}(TeV)
candidate has spin zero;
we will show that, even if it does not carry $J_R\geq 2$, the spectrum
must contain further particles that do.

If the possible new state is a scalar (or a pseudoscalar), $\mS$,
the sum-rule~(\ref{eq.SR-narrow}) becomes
\bear
16\pi  \, \Frac{\Gamma_{\mS\to \gamma\gamma}}{M_{\mS}^3}
&=&    \sum_{R\neq \mS} \, 16\pi \, (2J_R+1) \,
\Frac{ ( \Gamma_{R\to [\gamma\gamma]_2}   - \Gamma_{R\to [\gamma\gamma]_0} ) }{M_R^3}
\nn\\
&&\qquad \mbox{+ non-R.}
\label{eq.scalar-SR}
\eear
In order to fulfill the identity~(\ref{eq.scalar-SR}),
resonances with $\Gamma_{R\to [\gamma\gamma]_2}\neq 0$ are needed
on the right-hand side, i.e., resonances with spin $J_R\geq 2$.
We also briefly discuss the importance of non-R loop contributions below.

In the case of a  
resonance with $M_{\mS}\sim 1$~TeV
(the errors are already very large at this energy to allow for this possibility)
with a large $\gamma\gamma$ partial width 
($\sim 10$~GeV),  
the sum-rule~(\ref{eq.SR-narrow})~\cite{Csaki:2015vek}
gets the contribution
\bear
16\pi  \Frac{ \Gamma_{\mS\to \gamma\gamma}}{M_{\mS}^3}
\,\,\, \sim \,\,\,  0.5 \, \mbox{TeV}^{-2}\, .
\label{eq.S-contribution}
\eear
In comparison, the SM Higgs exchange yields the negligible contribution
$16\pi \Gamma_{h\to\gamma\gamma}/m_h^3  \simeq 2.4\times 10^{-4} \, \mbox{TeV}^{-2}$~\cite{pdg}.
This large difference indicates that, in order to have a scalar resonance with a loop-induced decay
like the SM one with similar scalar couplings,
one needs either a large number of particles running in the intermediate
loop or huge hypercharges~\cite{Franceschini:2015kwy}.~\footnote{See also Ref.~\cite{Csaki:2016raa},
where $\cO(1)$~GeV width is advocated in the case of production via $\gamma\gamma$ fusion.}

For radiatively generated $S\to \gamma\gamma$ decays  
in weakly interacting theories in the TeV,
small partial widths are expected.
For a Higgs-like decay $\mS\to \gamma\gamma$, one would have
$\Gamma_{\mS\to \gamma\gamma}\sim  \alpha^2 M_\mS^3/(256\pi^3 v^2) |\sum_i N_i Q_i^2 A_i|^2$,
summing over the number $N_i$ of particles $i$ running within the loop with electric charge $Q_i$ and
the $A_i$ coefficient from the
one-loop integral (e.g., $|A_{W}|\leq 12.4$ and $|A_t|\leq 2.4$)~\cite{Djouadi:2005gi}.
This yields a contribution to~(\ref{eq.SR-narrow}) of the order of
$16\pi  \Gamma_{\mS\to \gamma\gamma}/M_{\mS}^3 \sim  \alpha^2 /(16\pi^2 v^2) 
\sim  10^{-5}$~TeV$^{-2}$.

Background non-R diagrams also contribute to the sum-rule Eq.~(\ref{eq.SR}).
However, based on naive dimensional analysis, these
are found to be small,
of the order of
\bear
&&\Frac{1}{\pi}\Int_{\nu_{\rm th}}^\infty \, \, \Frac{\rm d\nu'}{(\nu')^{2}}  \,\,
{\rm Im}T_{\Delta\lambda}(\nu'+i\epsilon) \bigg|_{\rm non-R}
\nn\\
&&\qquad\qquad \qquad
  \sim  \Frac{\alpha^2}{\nu_{\rm th}}\quad\sim
\quad 10^{-4}\, \mbox{TeV}^{-2}\, ,
\eear
 where, for possible new physics states in the non-R loop
in an underlying weakly interacting theory (if any),
we expect the thresholds to be $\nu_{\rm th}> (750$~GeV$)^2$.
More precisely, in Quantum Electrodynamics (QED) with either a scalar or
a spin--$\frac{1}{2}$ particle with charge $Q=1$,
the $\gamma\gamma$ cross section difference
reaches a sharp global minimum with
$\sigma_2(\nu)-\sigma_0(\nu)
\gsim  - 8\alpha^2/\nu_{\rm th}$
right after the production threshold
due to the negative $\Delta\lambda=0$ contribution, then a wider global maximum
with              
$\sigma_2(\nu)-\sigma_0(\nu)  \lsim 2\alpha^2/\nu_{\rm th}$
due to positive $\Delta\lambda=2$ production,
and finally a converging $1/\nu$ tail~\cite{Pascalutsa:2012}.
Thus, one finds that the pure QED one-loop amplitude
for $\gamma\gamma\to\gamma\gamma$ fulfills the sum-rule~(\ref{eq.SR}) on
its own and yields no correction~\cite{Pascalutsa:2012}. Therefore,
in order to get a contribution from these background loops
to cancel the scalar resonance one
in Eq.~(\ref{eq.S-contribution}), one should incorporate effects
beyond QED, which first enter at two loops. We find it very unlikely
that these corrections are large enough to achieve this goal without
entering
a non-perturbative regime.

An elementary scalar that radiatively decays into two photons would not require in principle the addition of extra particles with $J_R\geq 2$,
beyond new fermions with huge charges --or a huge number of components-- if one considers a large diphoton partial width.
However, as we have shown, achieving the latter and the required large background non-resonant contribution to the sum-rule
implies a departure from perturbativity in the TeV range, where the BSM theory would enter a strongly coupled regime
(see, e.g., Ref.~\cite{Bertuzzo:2016fmv}).
Thus, one would expect to have composite states of any total angular momenta $J_R\geq 2$ lying in the non-perturbative energy range,
as nothing forbids excitations with an arbitrary orbital momentum, similar to what one observes in QCD.

We are then left with the need of incorporating further resonance contributions
with spin $J_R\geq 2$
to cancel out that from  
an $\cal{O}$(TeV)
scalar candidate in Eq.~(\ref{eq.S-contribution}).

Assuming that the sum-rule is dominated by the lowest spins and lightest resonances,
the following relation for the lightest tensor resonance $\mT$ is obtained, 
\bear \label{Simplest}
&&\Gamma_{\mT\to \gamma\gamma} \approx  \Frac{ \Gamma_{\mS\to \gamma\gamma} }{5} \left(\Frac{M_{\mT}}{M_{\mS}}\right)^3\,  , 
\label{eq.lowest-R-dominance}
\eear
where we take the case in which the tensor decays only
into $\gamma\gamma$ states with helicity difference $\Delta\lambda=2$.\footnote{
This is the case, for instance, for massive gravitons
$G$~\cite{RS,KK-graviton,Gouzevitch:2013qca},
the decay $G\to \gamma(\lambda)\gamma(\lambda')$ of which only occurs with $\Delta\lambda=2$.}
Otherwise, we would obtain a larger value of $\Gamma_{\mT\to \gamma\gamma}$ since
the $\Gamma_{\mT\to [\gamma\gamma]_0}$ would add up to the scalar contribution and
a larger $\Gamma_{\mT\to [\gamma\gamma]_2}$ would be needed to fulfill the sum-rule.

Finally, we note that these relations should eventually take into account additional
resonances in the TeV region, correcting our lowest-resonance dominance assumption.

\section{Conclusions}

The late 2015~\cite{exp} and early 2016~\cite{Moriond}  
analyses by the 
ATLAS and CMS collaborations
 caused a huge excitement in our
 community, as they might have been the first direct evidence of a particle beyond the SM.
 On the contrary, mid 2016 data~\cite{ICHEP16}
 did not show any significant diphoton excess. Of course, the diphoton channel continues to be one of the golden channels for new physics searches, which calls for a model-independent
 analysis of the consequences of a diphoton bump as we carried out in this paper.

 The application of axiomatic quantum field theory to
 the diphoton production allows us
to conclude that:\\
 i)       
  A diphoton signal with an $\cO($TeV$)$ mass
 confirmed as a broad spin-0 or 2 resonance   
 will not come alone; as other resonances with different spin are needed
 to satisfy the derived sum
 rules. According to our discussion, this is obvious for the spin-0 case.
 Crossing symmetry implies that spin-2 (or higher) exchanges violate partial-wave unitarity,
 requiring the existence of a
 Regge-like tower with higher spin states. Odd-spins (3, 5, ...) need
 contributions from even spins --most likely $J_R=0$--
 to satisfy the sum rule~(\ref{eq.SR-narrow}) and require a tower of
 higher spins to restore partial-wave unitarity.
 All the reasoning in this article assumes a large width, of the order of the GeV;
 for much narrower resonances,
 the sum-rule cancellation can be achieved via non-R loop diagrams.
 \\
 ii) As a secondary conclusion, we remark that
the proposed $\gamma\gamma$ sum-rules must be fulfilled for any value of the partial widths.
Thus, in the case these had values much below the GeV,
the tiny  
contribution from a resonance with $M_R\sim 1$~TeV
to the sum-rule would still have to be cancelled out in a very precise
way. However,     
in this case, it could be achieved by means of perturbative non-resonant
BSM   
loops, which might be seen as a non-resonant background excess.
Although it may be experimentally challenging, the direct evaluation of the sum rule from the data could give an 
estimate
of the deviation from the SM and the presence of new physics.

 In the absence of precise enough data, the application of general quantum field theory principles allows
 the derivation of model-independent
 relations (applied here to the diphoton channel) that can be helpful in understanding 
 potential new phenomena.
 For this reason, due to its simplicity and generality, the analysis shown in this letter is of high interest for the study of  
future diphoton searches  
at colliders.

\section*{Acknowledgements}

We thank Ver\'onica Sanz for valuable comments on our draft.
P.R. acknowledges funding from Conacyt, M\'exico, through Projects 296 (``Fronteras de la Ciencia''), 
No. 236394, No. 250628 (``Ciencia B\'asica'')
and SNI.
The work of J.J.S.C. was supported by the Spanish Ministry MINECO under Grant No. FPA2013-44773-P
and the Centro de Excelencia Severo Ochoa Programme under
Grant No. SEV-2012-0249. J.J.S.C. enjoyed the hospitality of Physics Department, Cinvestav, 
where part of this research was developed.

\end{document}